\documentstyle[12pt,epsfig,amstex] {article}
\begin{document}
\newcommand{\vl}{\begin{picture}(1,10)
\newcommand{\qqbar}{Z^{0} \to q \bar q}
\newcommand{\gaga}{\gamma \gamma}
\newcommand{\ee} {\mathrm{e}^+ \mathrm{e}^-}
\newcommand{\mm}{\mu^+ \mu^-}
\newcommand{\tt}{\tau^+ \tau^-}
\newcommand{\MZ}      {m_{\mathrm{Z}}}
\put(4.5,-5){\line(0,1){30}}\end{picture}}
\begin{titlepage}
\pagestyle{empty}
  
\vskip 2.0 cm
 
\begin{center}
{\large EUROPEAN LABORATORY FOR PARTICLE PHYSICS}
\end{center}

\vskip 1.0 cm

\begin{flushright}
CERN-PPE/97-041   \\
16 April 1997 \\
\vskip 0.5 cm 
\end{flushright}

\vskip 1.0 cm

\begin{center}
\boldmath  
{\LARGE\bf Search for 
pair-production of
long-lived heavy 
charged particles in $\mathrm{e}^+ \mathrm{e}^-$ annihilation}

\vskip 1.0 cm
{\LARGE The ALEPH Collaboration}
\end{center}
\vskip 3 cm

\centerline{\large\bf Abstract}
\vskip 0.6 cm
A search for pair-production of
long-lived, heavy, singly-charged particles 
has been performed with data collected 
by the ALEPH detector at a centre-of-mass energy 
of 172~GeV. Data at $\sqrt{s} = 161$, 136, and 130~GeV 
are also included to improve
the sensitivity to lower masses.
No candidate is found in the data.
A model-independent  95$\%$~confidence level
upper limit on the production cross section at 172~GeV 
of 0.2--0.4~pb is derived for masses between 45 and 86~GeV/$c^2$.
This cross section limit implies, assuming the MSSM, 
a lower limit of 67 (69) GeV/$c^2$ on the mass of 
right- (left-) handed long-lived scalar taus or scalar
muons and of 86 GeV/$c^2$ on the mass of long-lived
charginos.

\vskip 2.0 cm
\begin{center}
(submitted to Physics Letters B)
\end{center}

\vskip 1.5 cm
\vfill
\vbox{
\hrule width6cm height0.5pt\vskip4pt 
\noindent $^*$See the following pages for the list of authors.}

\end{titlepage}

\pagestyle{empty}
\newpage
\small
%
%
\newlength{\saveparskip}
\newlength{\savetextheight}
\newlength{\savetopmargin}
\newlength{\savetextwidth}
\newlength{\saveoddsidemargin}
\newlength{\savetopsep}
\setlength{\saveparskip}{\parskip}
\setlength{\savetextheight}{\textheight}
\setlength{\savetopmargin}{\topmargin}
\setlength{\savetextwidth}{\textwidth}
\setlength{\saveoddsidemargin}{\oddsidemargin}
\setlength{\savetopsep}{\topsep}
%
%
\setlength{\parskip}{0.0cm}
\setlength{\textheight}{25.0cm}
\setlength{\topmargin}{-1.5cm}
\setlength{\textwidth}{16 cm}
\setlength{\oddsidemargin}{-0.0cm}
\setlength{\topsep}{1mm}
\pretolerance=10000
\centerline{\large\bf The ALEPH Collaboration}
\footnotesize
\vspace{0.5cm}
{\raggedbottom
\begin{sloppypar}
\samepage\noindent
R.~Barate,
D.~Buskulic,
D.~Decamp,
P.~Ghez,
C.~Goy,
J.-P.~Lees,
A.~Lucotte,
M.-N.~Minard,
J.-Y.~Nief,
B.~Pietrzyk
\nopagebreak
\begin{center}
\parbox{15.5cm}{\sl\samepage
Laboratoire de Physique des Particules (LAPP), IN$^{2}$P$^{3}$-CNRS,
74019 Annecy-le-Vieux Cedex, France}
\end{center}\end{sloppypar}
\vspace{2mm}
\begin{sloppypar}
\noindent
M.P.~Casado,
M.~Chmeissani,
P.~Comas,
J.M.~Crespo,
M.~Delfino, 
E.~Fernandez,
M.~Fernandez-Bosman,
Ll.~Garrido,$^{15}$
A.~Juste,
M.~Martinez,
R.~Miquel,
Ll.M.~Mir,
S.~Orteu,
C.~Padilla,
I.C.~Park,
A.~Pascual,
J.A.~Perlas,
I.~Riu,
F.~Sanchez,
F.~Teubert
\nopagebreak
\begin{center}
\parbox{15.5cm}{\sl\samepage
Institut de F\'{i}sica d'Altes Energies, Universitat Aut\`{o}noma
de Barcelona, 08193 Bellaterra (Barcelona), Spain$^{7}$}
\end{center}\end{sloppypar}
\vspace{2mm}
\begin{sloppypar}
\noindent
A.~Colaleo,
D.~Creanza,
M.~de~Palma,
G.~Gelao,
G.~Iaselli,
G.~Maggi,
M.~Maggi,
N.~Marinelli,
S.~Nuzzo,
A.~Ranieri,
G.~Raso,
F.~Ruggieri,
G.~Selvaggi,
L.~Silvestris,
P.~Tempesta,
A.~Tricomi,$^{3}$
G.~Zito
\nopagebreak
\begin{center}
\parbox{15.5cm}{\sl\samepage
Dipartimento di Fisica, INFN Sezione di Bari, 70126
Bari, Italy}
\end{center}\end{sloppypar}
\vspace{2mm}
\begin{sloppypar}
\noindent
X.~Huang,
J.~Lin,
Q. Ouyang,
T.~Wang,
Y.~Xie,
R.~Xu,
S.~Xue,
J.~Zhang,
L.~Zhang,
W.~Zhao
\nopagebreak
\begin{center}
\parbox{15.5cm}{\sl\samepage
Institute of High-Energy Physics, Academia Sinica, Beijing, The People's
Republic of China$^{8}$}
\end{center}\end{sloppypar}
\vspace{2mm}
\begin{sloppypar}
\noindent
D.~Abbaneo,
R.~Alemany,
A.O.~Bazarko,$^{1}$
U.~Becker,
P.~Bright-Thomas,
M.~Cattaneo,
F.~Cerutti,
G.~Dissertori,
H.~Drevermann,
R.W.~Forty,
M.~Frank,
R.~Hagelberg,
J.B.~Hansen,
J.~Harvey,
P.~Janot,
B.~Jost,
E.~Kneringer,
J.~Knobloch,
I.~Lehraus,
G.~Lutters,
P.~Mato,
A.~Minten,
L.~Moneta,
A.~Pacheco,
J.-F.~Pusztaszeri,$^{21}$
F.~Ranjard,
G.~Rizzo,
L.~Rolandi,
D.~Schlatter,
M.~Schmitt,
O.~Schneider,
W.~Tejessy,
I.R.~Tomalin,
H.~Wachsmuth,
A.~Wagner
\nopagebreak
\begin{center}
\parbox{15.5cm}{\sl\samepage
European Laboratory for Particle Physics (CERN), 1211 Geneva 23,
Switzerland}
\end{center}\end{sloppypar}
\vspace{2mm}
\begin{sloppypar}
\noindent
Z.~Ajaltouni,
A.~Barr\`{e}s,
C.~Boyer,
A.~Falvard,
C.~Ferdi,
P.~Gay,
C~.~Guicheney,
P.~Henrard,
J.~Jousset,
B.~Michel,
S.~Monteil,
J-C.~Montret,
D.~Pallin,
P.~Perret,
F.~Podlyski,
J.~Proriol,
P.~Rosnet,
J.-M.~Rossignol
\nopagebreak
\begin{center}
\parbox{15.5cm}{\sl\samepage
Laboratoire de Physique Corpusculaire, Universit\'e Blaise Pascal,
IN$^{2}$P$^{3}$-CNRS, Clermont-Ferrand, 63177 Aubi\`{e}re, France}
\end{center}\end{sloppypar}
\vspace{2mm}
\begin{sloppypar}
\noindent
T.~Fearnley,
J.D.~Hansen,
J.R.~Hansen,
P.H.~Hansen,
B.S.~Nilsson,
B.~Rensch,
A.~W\"a\"an\"anen
\begin{center}
\parbox{15.5cm}{\sl\samepage
Niels Bohr Institute, 2100 Copenhagen, Denmark$^{9}$}
\end{center}\end{sloppypar}
\vspace{2mm}
\begin{sloppypar}
\noindent
G.~Daskalakis,
A.~Kyriakis,
C.~Markou,
E.~Simopoulou,
A.~Vayaki
\nopagebreak
\begin{center}
\parbox{15.5cm}{\sl\samepage
Nuclear Research Center Demokritos (NRCD), Athens, Greece}
\end{center}\end{sloppypar}
\vspace{2mm}
\begin{sloppypar}
\noindent
A.~Blondel,
J.C.~Brient,
F.~Machefert,
A.~Roug\'{e},
M.~Rumpf,
A.~Valassi,$^{6}$
H.~Videau
\nopagebreak
\begin{center}
\parbox{15.5cm}{\sl\samepage
Laboratoire de Physique Nucl\'eaire et des Hautes Energies, Ecole
Polytechnique, IN$^{2}$P$^{3}$-CNRS, 91128 Palaiseau Cedex, France}
\end{center}\end{sloppypar}
\vspace{2mm}
\begin{sloppypar}
\noindent
E.~Focardi,
G.~Parrini,
K.~Zachariadou
\nopagebreak
\begin{center}
\parbox{15.5cm}{\sl\samepage
Dipartimento di Fisica, Universit\`a di Firenze, INFN Sezione di Firenze,
50125 Firenze, Italy}
\end{center}\end{sloppypar}
\vspace{2mm}
\begin{sloppypar}
\noindent
R.~Cavanaugh,
M.~Corden,
C.~Georgiopoulos,
T.~Huehn,
D.E.~Jaffe
\nopagebreak
\begin{center}
\parbox{15.5cm}{\sl\samepage
Supercomputer Computations Research Institute,
Florida State University,
Tallahassee, FL 32306-4052, USA $^{13,14}$}
\end{center}\end{sloppypar}
\vspace{2mm}
\begin{sloppypar}
\noindent
A.~Antonelli,
G.~Bencivenni,
G.~Bologna,$^{4}$
F.~Bossi,
P.~Campana,
G.~Capon,
D.~Casper,
V.~Chiarella,
G.~Felici,
P.~Laurelli,
G.~Mannocchi,$^{5}$
F.~Murtas,
G.P.~Murtas,
L.~Passalacqua,
M.~Pepe-Altarelli
\nopagebreak
\begin{center}
\parbox{15.5cm}{\sl\samepage
Laboratori Nazionali dell'INFN (LNF-INFN), 00044 Frascati, Italy}
\end{center}\end{sloppypar}
\vspace{2mm}
\begin{sloppypar}
\noindent
L.~Curtis,
S.J.~Dorris,
A.W.~Halley,
I.G.~Knowles,
J.G.~Lynch,
V.~O'Shea,
C.~Raine,
J.M.~Scarr,
K.~Smith,
P.~Teixeira-Dias,
A.S.~Thompson,
E.~Thomson,
F.~Thomson,
R.M.~Turnbull
\nopagebreak
\begin{center}
\parbox{15.5cm}{\sl\samepage
Department of Physics and Astronomy, University of Glasgow, Glasgow G12
8QQ,United Kingdom$^{10}$}
\end{center}\end{sloppypar}
\vspace{2mm}
\begin{sloppypar}
\noindent
C.~Geweniger,
G.~Graefe,
P.~Hanke,
G.~Hansper,
V.~Hepp,
E.E.~Kluge,
A.~Putzer,
M.~Schmidt,
J.~Sommer,
K.~Tittel,
S.~Werner,
M.~Wunsch
\begin{center}
\parbox{15.5cm}{\sl\samepage
Institut f\"ur Hochenergiephysik, Universit\"at Heidelberg, 69120
Heidelberg, Fed.\ Rep.\ of Germany$^{16}$}
\end{center}\end{sloppypar}
\vspace{2mm}
\begin{sloppypar}
\noindent
R.~Beuselinck,
D.M.~Binnie,
W.~Cameron,
P.J.~Dornan,
M.~Girone,
S.~Goodsir,
E.B.~Martin,
P.~Morawitz,
A.~Moutoussi,
J.~Nash,
J.K.~Sedgbeer,
A.M.~Stacey,
M.D.~Williams
\nopagebreak
\begin{center}
\parbox{15.5cm}{\sl\samepage
Department of Physics, Imperial College, London SW7 2BZ,
United Kingdom$^{10}$}
\end{center}\end{sloppypar}
\vspace{2mm}
\begin{sloppypar}
\noindent
P.~Girtler,
D.~Kuhn,
G.~Rudolph
\nopagebreak
\begin{center}
\parbox{15.5cm}{\sl\samepage
Institut f\"ur Experimentalphysik, Universit\"at Innsbruck, 6020
Innsbruck, Austria$^{18}$}
\end{center}\end{sloppypar}
\vspace{2mm}
\begin{sloppypar}
\noindent
A.P.~Betteridge,
C.K.~Bowdery,
P.~Colrain,
G.~Crawford,
A.J.~Finch,
F.~Foster,
G.~Hughes,
R.W.~Jones,
T.~Sloan,
E.P.~Whelan,
M.I.~Williams
\nopagebreak
\begin{center}
\parbox{15.5cm}{\sl\samepage
Department of Physics, University of Lancaster, Lancaster LA1 4YB,
United Kingdom$^{10}$}
\end{center}\end{sloppypar}
\vspace{2mm}
\begin{sloppypar}
\noindent
C.~Hoffmann,
K.~Jakobs,
K.~Kleinknecht,
G.~Quast,
B.~Renk,
E.~Rohne,
H.-G.~Sander,
P.~van~Gemmeren,
C.~Zeitnitz
\nopagebreak
\begin{center}
\parbox{15.5cm}{\sl\samepage
Institut f\"ur Physik, Universit\"at Mainz, 55099 Mainz, Fed.\ Rep.\
of Germany$^{16}$}
\end{center}\end{sloppypar}
\vspace{2mm}
\begin{sloppypar}
\noindent
J.J.~Aubert,
C.~Benchouk,
A.~Bonissent,
G.~Bujosa,
D.~Calvet,
J.~Carr,
P.~Coyle,
C.~Diaconu,
N.~Konstantinidis,
O.~Leroy,
F.~Motsch,
P.~Payre,
D.~Rousseau,
M.~Talby,
A.~Sadouki,
M.~Thulasidas,
A.~Tilquin,
K.~Trabelsi
\nopagebreak
\begin{center}
\parbox{15.5cm}{\sl\samepage
Centre de Physique des Particules, Facult\'e des Sciences de Luminy,
IN$^{2}$P$^{3}$-CNRS, 13288 Marseille, France}
\end{center}\end{sloppypar}
\vspace{2mm}
\begin{sloppypar}
\noindent
M.~Aleppo, 
F.~Ragusa$^{12}$
\nopagebreak
\begin{center}
\parbox{15.5cm}{\sl\samepage
Dipartimento di Fisica, Universit\`a di Milano e INFN Sezione di
Milano, 20133 Milano, Italy.}
\end{center}\end{sloppypar}
\vspace{2mm}
\begin{sloppypar}
\noindent
R.~Berlich,
W.~Blum,
V.~B\"uscher,
H.~Dietl,
G.~Ganis,
C.~Gotzhein,
H.~Kroha,
G.~L\"utjens,
G.~Lutz,
W.~M\"anner,
H.-G.~Moser,
R.~Richter,
A.~Rosado-Schlosser,
S.~Schael,
R.~Settles,
H.~Seywerd,
R.~St.~Denis,
H.~Stenzel,
W.~Wiedenmann,
G.~Wolf
\nopagebreak
\begin{center}
\parbox{15.5cm}{\sl\samepage
Max-Planck-Institut f\"ur Physik, Werner-Heisenberg-Institut,
80805 M\"unchen, Fed.\ Rep.\ of Germany\footnotemark[16]}
\end{center}\end{sloppypar}
\vspace{2mm}
\begin{sloppypar}
\noindent
J.~Boucrot,
O.~Callot,$^{12}$
S.~Chen,
A.~Cordier,
M.~Davier,
L.~Duflot,
J.-F.~Grivaz,
Ph.~Heusse,
A.~H\"ocker,
A.~Jacholkowska,
M.~Jacquet,
D.W.~Kim,$^{2}$
F.~Le~Diberder,
J.~Lefran\c{c}ois,
A.-M.~Lutz,
I.~Nikolic,
M.-H.~Schune,
S.~Simion,
E.~Tournefier,
J.-J.~Veillet,
I.~Videau,
D.~Zerwas
\nopagebreak
\begin{center}
\parbox{15.5cm}{\sl\samepage
Laboratoire de l'Acc\'el\'erateur Lin\'eaire, Universit\'e de Paris-Sud,
IN$^{2}$P$^{3}$-CNRS, 91405 Orsay Cedex, France}
\end{center}\end{sloppypar}
\vspace{2mm}
\begin{sloppypar}
\noindent
\samepage
P.~Azzurri,
G.~Bagliesi,
G.~Batignani,
S.~Bettarini,
C.~Bozzi,
G.~Calderini,
M.~Carpinelli,
M.A.~Ciocci,
V.~Ciulli,
R.~Dell'Orso,
R.~Fantechi,
I.~Ferrante,
A.~Giassi,
A.~Gregorio,
F.~Ligabue,
A.~Lusiani,
P.S.~Marrocchesi,
A.~Messineo,
F.~Palla,
G.~Sanguinetti,
A.~Sciab\`a,
P.~Spagnolo,
J.~Steinberger,
R.~Tenchini,
G.~Tonelli,$^{20}$
C.~Vannini,
A.~Venturi,
P.G.~Verdini
\samepage
\begin{center}
\parbox{15.5cm}{\sl\samepage
Dipartimento di Fisica dell'Universit\`a, INFN Sezione di Pisa,
e Scuola Normale Superiore, 56010 Pisa, Italy}
\end{center}\end{sloppypar}
\vspace{2mm}
\begin{sloppypar}
\noindent
G.A.~Blair,
L.M.~Bryant,
J.T.~Chambers,
Y.~Gao,
M.G.~Green,
T.~Medcalf,
P.~Perrodo,
J.A.~Strong,
J.H.~von~Wimmersperg-Toeller
\nopagebreak
\begin{center}
\parbox{15.5cm}{\sl\samepage
Department of Physics, Royal Holloway \& Bedford New College,
University of London, Surrey TW20 OEX, United Kingdom$^{10}$}
\end{center}\end{sloppypar}
\vspace{2mm}
\begin{sloppypar}
\noindent
D.R.~Botterill,
R.W.~Clifft,
T.R.~Edgecock,
S.~Haywood,
P.~Maley,
P.R.~Norton,
J.C.~Thompson,
A.E.~Wright
\nopagebreak
\begin{center}
\parbox{15.5cm}{\sl\samepage
Particle Physics Dept., Rutherford Appleton Laboratory,
Chilton, Didcot, Oxon OX11 OQX, United Kingdom$^{10}$}
\end{center}\end{sloppypar}
\vspace{2mm}
\begin{sloppypar}
\noindent
B.~Bloch-Devaux,
P.~Colas,
B.~Fabbro,
W.~Kozanecki,
E.~Lan\c{c}on,
M.C.~Lemaire,
E.~Locci,
P.~Perez,
J.~Rander,
J.-F.~Renardy,
A.~Rosowsky,
A.~Roussarie,
J.-P.~Schuller,
J.~Schwindling,
A.~Trabelsi,
B.~Vallage
\nopagebreak
\begin{center}
\parbox{15.5cm}{\sl\samepage
CEA, DAPNIA/Service de Physique des Particules,
CE-Saclay, 91191 Gif-sur-Yvette Cedex, France$^{17}$}
\end{center}\end{sloppypar}
\vspace{2mm}
\begin{sloppypar}
\noindent
S.N.~Black,
J.H.~Dann,
H.Y.~Kim,
A.M.~Litke,
M.A. McNeil,
G.~Taylor
\nopagebreak
\begin{center}
\parbox{15.5cm}{\sl\samepage
Institute for Particle Physics, University of California at
Santa Cruz, Santa Cruz, CA 95064, USA$^{19}$}
\end{center}\end{sloppypar}
\vspace{2mm}
\begin{sloppypar}
\noindent
C.N.~Booth,
R.~Boswell,
C.A.J.~Brew,
S.~Cartwright,
F.~Combley,
M.S.~Kelly,
M.~Lehto,
W.M.~Newton,
J.~Reeve,
L.F.~Thompson
\nopagebreak
\begin{center}
\parbox{15.5cm}{\sl\samepage
Department of Physics, University of Sheffield, Sheffield S3 7RH,
United Kingdom$^{10}$}
\end{center}\end{sloppypar}
\vspace{2mm}
\begin{sloppypar}
\noindent
K.~Affholderbach,
A.~B\"ohrer,
S.~Brandt,
G.~Cowan,
J.~Foss,
C.~Grupen,
P.~Saraiva,
L.~Smolik,
F.~Stephan 
\nopagebreak
\begin{center}
\parbox{15.5cm}{\sl\samepage
Fachbereich Physik, Universit\"at Siegen, 57068 Siegen,
 Fed.\ Rep.\ of Germany$^{16}$}
\end{center}\end{sloppypar}
\vspace{2mm}
\begin{sloppypar}
\noindent
M.~Apollonio,
L.~Bosisio,
R.~Della~Marina,
G.~Giannini,
B.~Gobbo,
G.~Musolino
\nopagebreak
\begin{center}
\parbox{15.5cm}{\sl\samepage
Dipartimento di Fisica, Universit\`a di Trieste e INFN Sezione di Trieste,
34127 Trieste, Italy}
\end{center}\end{sloppypar}
\vspace{2mm}
\begin{sloppypar}
\noindent
J.~Putz,
J.~Rothberg,
S.~Wasserbaech,
R.W.~Williams
\nopagebreak
\begin{center}
\parbox{15.5cm}{\sl\samepage
Experimental Elementary Particle Physics, University of Washington, WA 98195
Seattle, U.S.A.}
\end{center}\end{sloppypar}
\vspace{2mm}
\begin{sloppypar}
\noindent
S.R.~Armstrong,
E.~Charles,
P.~Elmer,
D.P.S.~Ferguson,
S.~Gonz\'{a}lez,
T.C.~Greening,
O.J.~Hayes,
H.~Hu,
S.~Jin,
P.A.~McNamara III,
J.M.~Nachtman,
J.~Nielsen,
W.~Orejudos,
Y.B.~Pan,
Y.~Saadi,
I.J.~Scott,
J.~Walsh,
Sau~Lan~Wu,
X.~Wu,
J.M.~Yamartino,
G.~Zobernig
\nopagebreak
\begin{center}
\parbox{15.5cm}{\sl\samepage
Department of Physics, University of Wisconsin, Madison, WI 53706,
USA$^{11}$}
\end{center}\end{sloppypar}
}
\footnotetext[1]{Now at Princeton University, Princeton, NJ 08544, U.S.A.}
\footnotetext[2]{Permanent address: Kangnung National University, Kangnung,
Korea.}
\footnotetext[3]{Also at Dipartimento di Fisica, INFN Sezione di Catania,
Catania, Italy.}
\footnotetext[4]{Also Istituto di Fisica Generale, Universit\`{a} di
Torino, Torino, Italy.}
\footnotetext[5]{Also Istituto di Cosmo-Geofisica del C.N.R., Torino,
Italy.}
\footnotetext[6]{Supported by the Commission of the European Communities,
contract ERBCHBICT941234.}
\footnotetext[7]{Supported by CICYT, Spain.}
\footnotetext[8]{Supported by the National Science Foundation of China.}
\footnotetext[9]{Supported by the Danish Natural Science Research Council.}
\footnotetext[10]{Supported by the UK Particle Physics and Astronomy Research
Council.}
\footnotetext[11]{Supported by the US Department of Energy, grant
DE-FG0295-ER40896.}
\footnotetext[12]{Also at CERN, 1211 Geneva 23,Switzerland.}
\footnotetext[13]{Supported by the US Department of Energy, contract
DE-FG05-92ER40742.}
\footnotetext[14]{Supported by the US Department of Energy, contract
DE-FC05-85ER250000.}
\footnotetext[15]{Permanent address: Universitat de Barcelona, 08208 Barcelona,
Spain.}
\footnotetext[16]{Supported by the Bundesministerium f\"ur Bildung,
Wissenschaft, Forschung und Technologie, Fed. Rep. of Germany.}
\footnotetext[17]{Supported by the Direction des Sciences de la
Mati\`ere, C.E.A.}
\footnotetext[18]{Supported by Fonds zur F\"orderung der wissenschaftlichen
Forschung, Austria.}
\footnotetext[19]{Supported by the US Department of Energy,
grant DE-FG03-92ER40689.}
\footnotetext[20]{Also at Istituto di Matematica e Fisica,
Universit\`a di Sassari, Sassari, Italy.}
\footnotetext[21]{Now at School of Operations Research and Industrial
Engireering, Cornell University, Ithaca, NY 14853-3801, U.S.A.}
%
%
\setlength{\parskip}{\saveparskip}
\setlength{\textheight}{\savetextheight}
\setlength{\topmargin}{\savetopmargin}
\setlength{\textwidth}{\savetextwidth}
\setlength{\oddsidemargin}{\saveoddsidemargin}
\setlength{\topsep}{\savetopsep}
\normalsize
\newpage
\pagestyle{plain}
\setcounter{page}{1}
\setcounter{footnote}{0}

\section{Introduction}

Most searches for supersymmetric (SUSY)~\cite{SUSY} particles 
assume that the lightest supersymmetric particle (LSP) 
is neutral and weakly interacting~\cite{yellow}
and that all charged SUSY particles ultimately decay, with a
very small lifetime, into the LSP 
plus ``standard" particles. 
In the hypothesis
of R-parity conservation, 
this suggests missing energy as a possible signature of SUSY.

Nevertheless, the possibility of 
supersymmetric long-lived charged particles which are not strongly 
interacting is favoured by some interesting classes of models. 
In Ref.~\cite{Dimopoulos}, for example,
a scenario is proposed in which the slepton is the 
next-to-lightest supersymmetric particle (NLSP). The slepton 
can decay into a standard lepton plus a Goldstino 
($\tilde{\ell} \to \ell \tilde{G}$) 
with a lifetime dependent on the 
SUSY-breaking energy scale $\sqrt{F}$:
$$c\tau \simeq (130~\mu{\mathrm{m}}) \left(\frac{100~{\mathrm{GeV}}/c^{2}}
{ m_{\tilde{\ell}}}  
\right)^5
 \left(\frac{\sqrt{F}}{100~{\mathrm{TeV}}}\right)^4 .$$ Hence at
LEP 2 centre-of-mass energies, if $\sqrt{F}$ is larger than a few thousand TeV,
a~70 GeV/$c^2$ slepton has a decay length greater than 10 meters, which 
corresponds to the typical size of the detectors of the LEP experiments.

Even in the minimal supersymmetric model (MSSM) a mass degeneracy of a few 
hundred MeV/$c^2$ or less
between the lightest chargino and the lightest neutralino could also induce
a chargino lifetime long enough for the particle to decay outside the detector.
This degeneracy may arise in some regions of the MSSM parameter
space when relaxing the gauge unification condition ($M^{'} =  \frac{5}{3} M
\tan^{2}{\theta_{W} }$ where $M^{'}$ and $M$ are the bino and wino
masses).

In R-parity violating SUSY models the possibility of long-lived charged
weakly interacting particles also exists. If, for example, the slepton is the LSP
a single dominant baryon-number violating coupling via the operator $\bar U
\bar D \bar D$ \cite{rpv} would give a long lifetime to the slepton.

At LEP~1, searches for heavy stable charged particles were 
performed by DELPHI~\cite{DELPHIs}, OPAL~\cite{OPALs} and  
ALEPH~\cite{ALEPHs}, which sets
the most stringent LEP 1 upper limit
on the production cross section, 
roughly 1.5~pb at 95\% CL for masses between 34 and 44~GeV/$c^2$.
DELPHI~\cite{DELPHIs2} has analysed data
collected at energies up to $\sqrt{s} = 172$~GeV,
setting a limit of $\sim$ 0.3--0.5 ~pb for masses between 45 and 85~GeV/$c^2$. 

This letter presents the results of a search for pair-production
of long-lived, singly-charged, not strongly interacting particles with mass 
greater than  45~GeV/$c^2$.
The analysis is performed 
using data collected during the second 1996 LEP running period, at 
$\sqrt{s} = 172$~GeV.
To increase the sensitivity for low masses,
data collected at 161~GeV (the first 1996 running period)
and 130, 136~GeV (the 1995 high-energy run) are also
analysed.

The selection is based on kinematic criteria related to
the pair-production hypothesis, and on the specific 
energy-loss measurement ($\it dE/dx$), a powerful tool for exploring
the high-mass region. 
When the mass of the heavy particles approaches the kinematic limit, the 
energy loss becomes high enough to saturate the main tracking detector's electronics.
To recover this interesting mass region, a selection based on the search for 
saturated signals in the tracking detector has been developed.

Limits on the  cross section 
are translated into lower limits on the masses 
of long-lived slepton and 
charginos based on
the production cross sections predicted by the MSSM.

\section{The ALEPH detector}

A detailed description of the ALEPH detector can be found in Ref.~\cite{Alnim},
and an account of its performance as well as a description of the
standard analysis algorithms in Ref.~\cite{Alperf}. 
Only a brief overview is given here.

Charged particles are detected in the central part of the detector
consisting of a silicon vertex detector, a drift chamber (ITC)
and a time projection chamber (TPC), 
all immersed in a 
1.5~T axial magnetic field provided by a superconducting solenoidal coil.
A $1/p_{\mathrm{T}}$ resolution of $6\times10^{-4}({\rm GeV}/c)^{-1}$ is measured.
 
The TPC sense wires provide up to 338 measurements of the 
specific ionization, $\it dE/dx$, for each track. 
To ensure a reliable $\it dE/dx$ measurement, 
tracks in this analysis are required to have at least 50 associated wire hits.

Between the TPC and the coil, an electromagnetic
calorimeter (ECAL) is used to identify electrons and photons and to
measure their energy, 
complemented by luminosity calorimeters (LCAL and SICAL) in the small 
polar angle region.
The iron return yoke is instrumented  to provide
a measurement of the hadronic energy (HCAL) 
and, together with external
chambers, muon identification.

The two main ALEPH triggers relevant for this analysis are based on the 
coincidence between a track
candidate in the ITC and an energy deposit in the ECAL or HCAL modules
to which the track is pointing.

\section{Data and Monte Carlo samples}

Data collected at $\sqrt{s} = 172$\footnote{The 172 GeV sample includes 
1.1 pb$^{-1}$ collected at 170 GeV.}, 161, 136, and 130~GeV are analysed,
corresponding to integrated luminosities of 
10.6, 11.1, 2.9, and 2.9~pb$^{-1}$, respectively. 

All major background reactions 
are generated at 161 and
172~GeV using the full detector simulation.  
These include the annihilation processes
$\mathrm{e}^+ \mathrm{e}^- \to \mathrm{f} \bar{\mathrm{f}} (\gamma)$ and 
the various processes leading  to four-fermion final states  
($\mathrm{e}^+ \mathrm{e}^- \to \mathrm{W^+ W^-}$, 
$\mathrm{e}^+ \mathrm{e}^- \to \mathrm{We}\nu$, 
$\mathrm{e}^+ \mathrm{e}^- \to \mathrm{Ze^+ e^-}$ and 
$\mathrm{e}^+ \mathrm{e}^- \to \mathrm{Z\gamma^*}$).
These samples correspond to more than 100 times the
integrated luminosity of the data.  Two-photon processes
($\gamma\gamma\rightarrow \ell^+ \ell^-$ and 
$\gamma\gamma\rightarrow \mathrm{q} \bar{\mathrm{q}}$) 
are also simulated with an integrated luminosity about three times 
that of the data.
 
For the signal, pair production of stable singly-charged particles 
is simulated for masses between 45 and 86~GeV/$c^2$.
These particles are treated as heavy muons by the detector simulation 
program.

\section{Low- and intermediate-mass selections}

A preselection is applied to reject topologies which are 
clearly incompatible with pair-production of back-to-back, massive, long-lived particles.
Exactly two 
tracks are required,
both of which must be well reconstructed in the TPC.
Both tracks must lie away from the beam axis ($|\cos{\theta}| < 0.90$),
have at least one hit in the ITC,
and satisfy the conditions $|d_{01}|+|d_{02}| < 0.3$~cm and  
$|z_{01}|+|z_{02}| < 5$~cm, where $d_0$ ($z_0$) is the distance of closest 
approach to the beam axis 
in the transverse plane (to the interaction 
point along the beam direction).
The two tracks must have 
transverse momenta with respect to the beam axis, $p_{\rm T}$, greater than  
$0.1 \sqrt{s}$,
equal momenta within three times the estimated error derived from the track fit,
and the angle between them (acollinearity) must be  greater than 160$^\circ$. 
Both tracks must fail the electron identification cut~\cite{Alperf}, 
and deposit less than 20~GeV (50~GeV) in the ECAL (HCAL).
An event is rejected if it contains a photon
with energy above 250~MeV or if energy deposits are detected 
in the luminosity calorimeters.
 
The acollinearity cut rejects radiative Z returns and
two-photon processes.
The cuts on $p_{\mathrm{T}}$ and the equal momentum requirement suppress
two-photon and $\tau^+ \tau^-$ backgrounds.
The photon veto rejects radiative Z returns
and Bhabha events. 
The cuts on electromagnetic and hadronic energy
further improve Bhabha rejection,
the latter being relevant when the electrons enter an ECAL 
insensitive region. 

After this preselection, a large $\mu^+ \mu^-$ background still survives,
since the large mass of the signal has not yet been exploited.
Two further sets of selection criteria, based on 
the measured particle masses $m_{1,2}$, defined as 
$m_{i}=\sqrt{E_{\mathrm{beam}}^2 - p_{i}^2}$,
are therefore introduced. 

The so-called low-mass selection requires the two particles to have  masses 
in the range $0.52 < m_{1,2}/E_{\mathrm{beam}} < 0.80$ and an acollinearity 
greater than $174^\circ$.
In this mass range, the momenta of the tracks are high enough to produce an
energy loss similar to that of ordinary particles; therefore
$dE/dx$ information is not used.

The intermediate-mass selection requires masses in the range 
$0.80 < m_{1,2}/E_{\mathrm{beam}} < 0.98$ and a specific ionization 
$(R_{\mu 1} + R_{\mu 2}) \ge 10$.
Here $R_{\mu}$ is an estimator calculated by
comparing the measured $dE/dx$, $I$, to that expected for a muon 
$\langle I_{\mu} \rangle$:
$R_{\mu} = \left( {I-\langle I_{\mu} \rangle} \right) / {\sigma_{I}} $, 
where $\sigma_{I}$ is
the expected resolution of the measurement.
In order that a single track saturating the TPC electronics
does not suffice for the event to be selected,
this estimator is assigned a value of 5 when saturation  prevents
a direct $dE/dx$ measurement.

For events with both masses in the range 90--98$\%$ of the beam energy 
the equal momentum requirement is relaxed, since
the expected background (mainly $\rm{e^+ e^-} \to \tau^+ \tau^-$) 
is lower in this mass region.

The only backgrounds
surviving these cuts originate from the di-lepton channels,
with cross sections of 
10.4~fb for $\mu^+ \mu^-$ and 1.2~fb 
for $\tau^+ \tau^-$. This corresponds to an expected background of
0.3 events in the full  data sample.  

\section{High-mass selection}

The extension of the search to masses approaching the kinematic limit
($m/E_{\mathrm{beam}} \simeq 1$) requires a somewhat different approach than
the low- and intermediate-mass analyses.  
Particles with masses within
a few percent of the kinematic limit ionise
so heavily that the TPC digitising electronics saturate and
neither high-resolution spatial coordinates nor reliable $dE/dx$ measurements
are available. Saturation occurs at
approximately 25 times minimum-ionization. Saturated hits are not used by the standard
ALEPH tracking since they have a resolution much worse than
typical, unsaturated data.  Nevertheless, the presence of many saturated
channels is itself a distinctive feature of a high-mass
signal and in a clean two-track topology the characteristic
pattern of a charged particle helix can be easily recognised from
the spatial distribution of saturated hits. 

The ITC amplifiers do not saturate, but 
the limited resolution in the $z$ coordinate and the short 
lever-arm afforded by the ITC
do not allow a kinematic selection based on tracks reconstructed 
with this detector alone. The high-mass selection requires exactly two 
tracks reconstructed in the ITC, which may or may not have (unsaturated)
TPC hits associated to them.  Events with energy deposited in the luminosity
detectors are rejected.  Energy deposits in the ECAL and HCAL are used to
provide a three-dimensional point on each track candidate and verify
a roughly back-to-back topology. 
The two most energetic calorimeter objects are selected;
they must have an acollinearity of at least $160^{\circ}$,
opposite polar angles within $8.5^{\circ}$, and each contain
at least 50\% of the calorimetric energy deposited on their side 
of the detector (defined by a plane containing the interaction point and 
perpendicular to the beam direction).
Since low-$\beta$,
massive particles should not shower in the calorimeters,
the energy of these objects must be less than 5~GeV.

The positions of the two selected calorimeter objects, used as estimates
of where the particles may have exited the tracking volume, and the position 
of the interaction point, used as an estimate of 
the origin of these particles,
provide enough information for a single-helix fit of the transverse
momentum and polar angle of the two track candidates.  
The reconstructed helix must cross at least four TPC pad rows
on each side of the detector.  The particle masses $m_{1,2}$ 
must be between 90 and 99.5\% of the beam energy.

The estimated particle trajectories from the single-helix fit do not
rely on any information from the tracking detectors, so the pattern of all hits
(saturated and unsaturated) in the ITC and TPC can be checked for consistency.
An approximatively 5 standard deviations road is defined around the single-helix 
trajectory; 
this road has a width of 1~cm (in $r\phi$) for the ITC, and 5~cm 
(in both $r\phi$ and $z$) for
the TPC.  
At least 60\% of the ITC layers and 40\% of the
TPC pad rows crossed by the road must contain a hit; this confirms the
presence of charged tracks where predicted by the fit to 
calorimeter objects.  The
fairly low occupancy requirement in the TPC allows a track which projects onto
a crack between outer sectors to be retained if every pad row in the inner
sectors contains a hit.  

Finally, the high mass of both particles is confirmed by the $dE/dx$
measurements (if available) or by saturation of the TPC wires or
pads (if not) by requiring either $R_{\mu 1} + R_{\mu 2} \ge 10$ or
$f_{\mathrm{sat}} > 0.33$; $f_{\mathrm{sat}}$ is the fraction of TPC pad rows crossed by the fitted helix
which contain a saturated coordinate within the 5~cm road.

No background from the simulation survives the high-mass selection.

\section{Efficiency}
\label{sec:eff}

\begin{figure}
\begin{center}\mbox{\epsfig{file=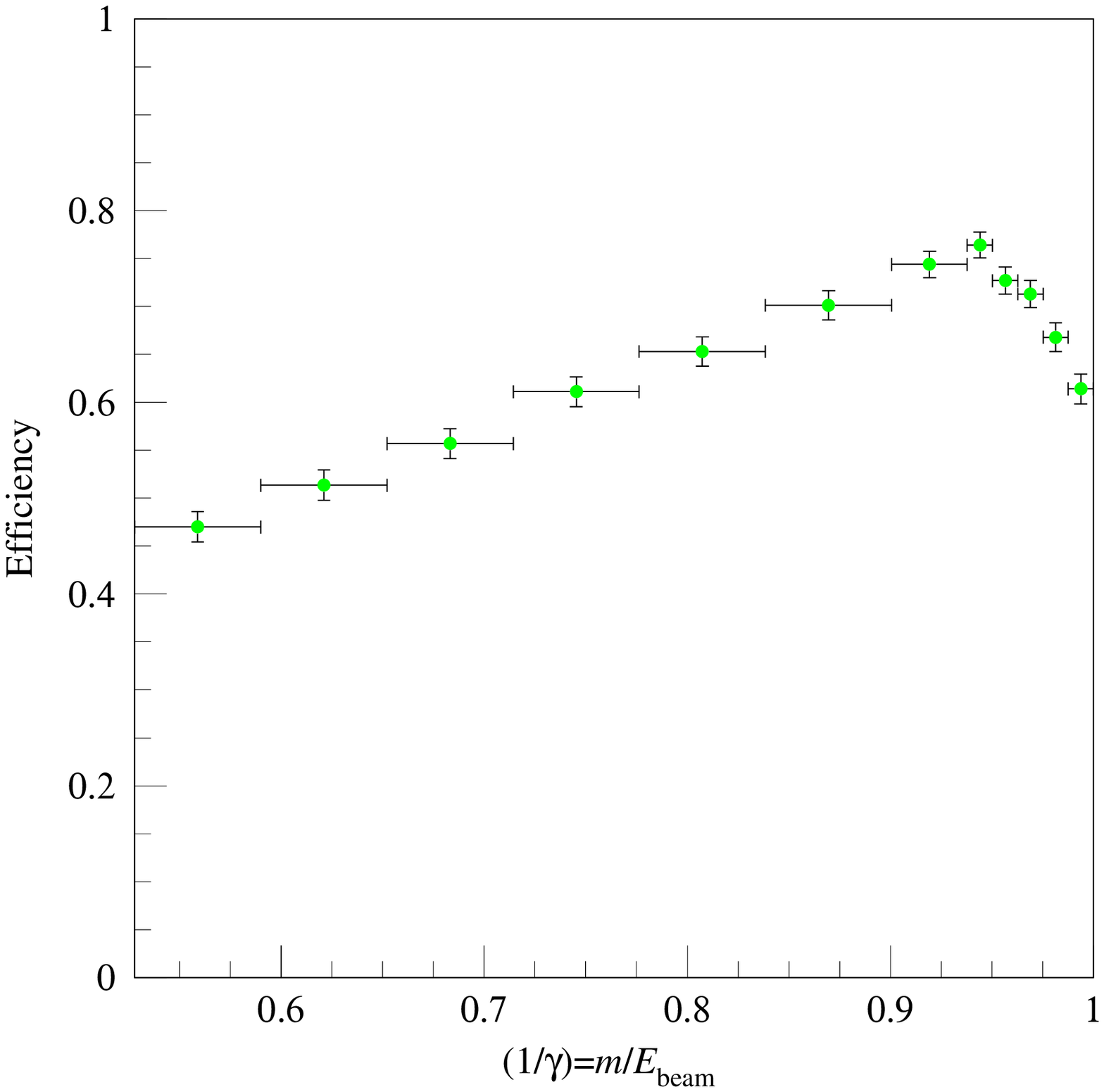,
height=15.cm}}
\end{center}
\caption{Selection efficiency as a function of $m/E_{\mathrm{beam}}$
for spin-1/2 stable charged particles.
The efficiency curve for spin-0 particles has a similar shape but is slightly
higher, owing to the larger fraction of cross section within the 
geometrical acceptance of the selection.}
\label{figura_eff}
\end{figure}

The final selection is a combination of the low- and
intermediate-mass criteria (including their common preselection),
and the complementary high-mass analysis.  Together, the three selections
yield an efficiency rising gently from 50\% at $m/E_{\mathrm{beam}}=0.50$ to
70\% at $m/E_{\mathrm{beam}} =0.90$;
this is shown in Fig.~\ref{figura_eff} 
where the selection efficiency (including the trigger) for spin-1/2 stable
charged particles is plotted as a function of  $m/E_{\mathrm{beam}}$.

For $m/E_{\mathrm{beam}} > 0.993$ the particles stop in the calorimeters 
because of the high energy loss. 
In this case, if the particles are not stable
and decay in the calorimeters, they might produce additional energy
deposits which will affect the selection efficiency.
In the Goldstino scenario (long-lived sleptons) 
the energy release can be important and the efficiency is conservatively assumed to be zero. 
However in the high mass-degeneracy scenario (long-lived charginos) almost all
the energy is taken by the neutralino and only a few hundred MeV are deposited
in the calorimeters. In this case the efficiency is assumed to be equal to 
that of stable particles. 

The efficiency depends on the spin (i.e., the 
production angular distribution)  
and mass of the charged particle. In the Monte Carlo signal generation,
massive particles are produced with a flat angular distribution. 
The spin-1/2 and spin-0 efficiencies are derived from the generated
signal by rescaling the angular acceptance using the formulae
\begin{eqnarray}\frac{d\sigma(\rm{spin}=0)}{d\Omega} &\propto&  
\frac{\beta^3}{s} \sin^{2}{\theta} \nonumber \\ 
\frac{d\sigma(\rm{spin}=1/2)}{d\Omega} &\propto&  
\frac{\beta}{s} \left[1 + \cos^{2}{\theta} + 
(1 - \beta^2)  \sin^{2}{\theta} \right] \ .
\label{css}
\end{eqnarray} 
The expression for the differential spin-1/2 cross section 
is valid for $s$ channel production under the assumption of purely vector coupling
to spin-1 bosons. 

To account for the dependence on $\sqrt{s}$,
all cuts are performed in terms of 
variables normalised to the beam energy 
and the efficiency is parametrized as a function of 
$m/E_{\mathrm{beam}}$.
The trigger efficiency has been estimated with a Monte Carlo program simulating 
the two ALEPH triggers briefly described in Section 2.
The trigger efficiency exceeds 99$\%$ for $m/E_{\mathrm{beam}} \le 0.97$, 
crosses 99$\%$ at $m/E_{\mathrm{beam}} \simeq 0.98$, and finally decreases to
80$\%$ for $m/E_{\mathrm{beam}} > 0.99$.

\section{Systematic checks}

For all variables used in the selections,
good agreement exists between
data and Monte Carlo simulation.
The most important variables used to isolate the signal events are 
the measured momenta and $\it dE/dx$ of the two particles.
The uncertainty on the beam energy has a negligible impact on the results.
The tracking performance was tested with data at the Z peak with a
variety of techniques (see for example~\cite{nu3}), showing agreement of
the absolute momentum calibration
between data and Monte Carlo better than $\sim 100$~MeV/$c$ for 45~GeV/$c$ tracks.
This level of momentum uncertainty has a negligible impact on the analysis.

Studies of $dE/dx$ performance at the Z peak have been described in 
many ALEPH papers (see for example ~\cite{nc}). 
Additional studies were performed with the 1996 data,
using minimum-ionising pions, identified muons, and electrons
identified using the ECAL information.
Agreement of the absolute $dE/dx$ calibration 
is obtained between data and Monte Carlo
at the level of $\sim 0.3$~ times the expected 
resolution, which is $\sim 4.5 \%$ for high energy 
isolated electrons.
The 1996 data exhibit a $dE/dx$ resolution 
\mbox{10--$20\%$} better than the Monte Carlo which goes in the direction of 
increasing the selection efficiency.
These effects lead to a very small change ($<1\%$) 
of the selection efficiency
since the $dE/dx$ information is only used in a
$\beta$ region when the 
expected ionization is much larger than the cut
applied in the analysis.

For the high-mass selection, the identification of pair-produced charged tracks
using calorimeter objects has been tested using di-muon
events collected at the Z peak.  
Although muons produced in Z decays  do not ionise heavily enough 
to saturate the TPC electronics, the high-mass signal can be simulated
with data by applying to di-muon hits the same algorithm
used for saturated coordinates.  
The resulting TPC hit positions agree 
with those of the standard ALEPH reconstruction
to within about half the width of a cluster on the pad row in $r\phi$
($\sim 1$~cm) or half the pulse-length in $z$ ($\ll 1$~cm).

The transverse spatial resolution of the ECAL entry point used in
the fit is roughly 1.5~cm, consistent with the granularity of the ECAL
towers.
The longitudinal uncertainty on the entry point
is expected to be of the order of one radiation length in lead, 
or less than 2~cm, and in
any case has only a second-order effect on the fit.  

A conservative 5$\%$ systematic error on the selection efficiency has been
applied. The impact of this 
systematic error on the cross section upper limit has been 
calculated according to the method described in Ref.~\cite{cousins},
based on the convolution of the Poisson probability estimator
with a Gaussian of sigma equal to the error on the signal efficiency.

\section{Results}

\begin{figure}
\begin{center}\mbox{\epsfig{file=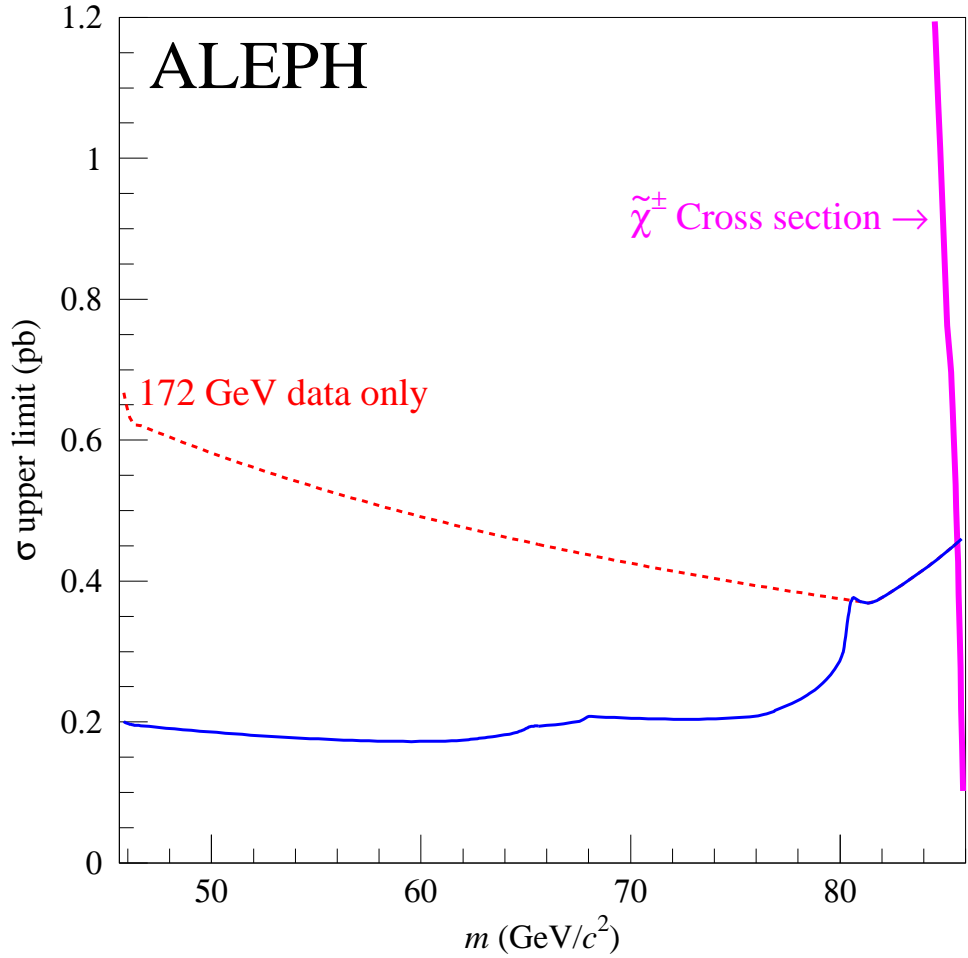,
height=15.cm}}
\end{center}
\caption{95$\%$ CL upper limit on the production cross section
for spin-1/2 particles at $\sqrt{s}=172$~GeV, as a function of mass.
The dashed curve shows the limit obtained from the 172~GeV data set alone, 
while the thick solid curve shows the limit obtained by
adding the 161, 136 and 130~GeV data to the analysis. 
The band on the right shows the chargino cross section 
calculated as described in Section 8.}
\label{figura1}
\end{figure}

\begin{figure}
\begin{center}\mbox{\epsfig{file=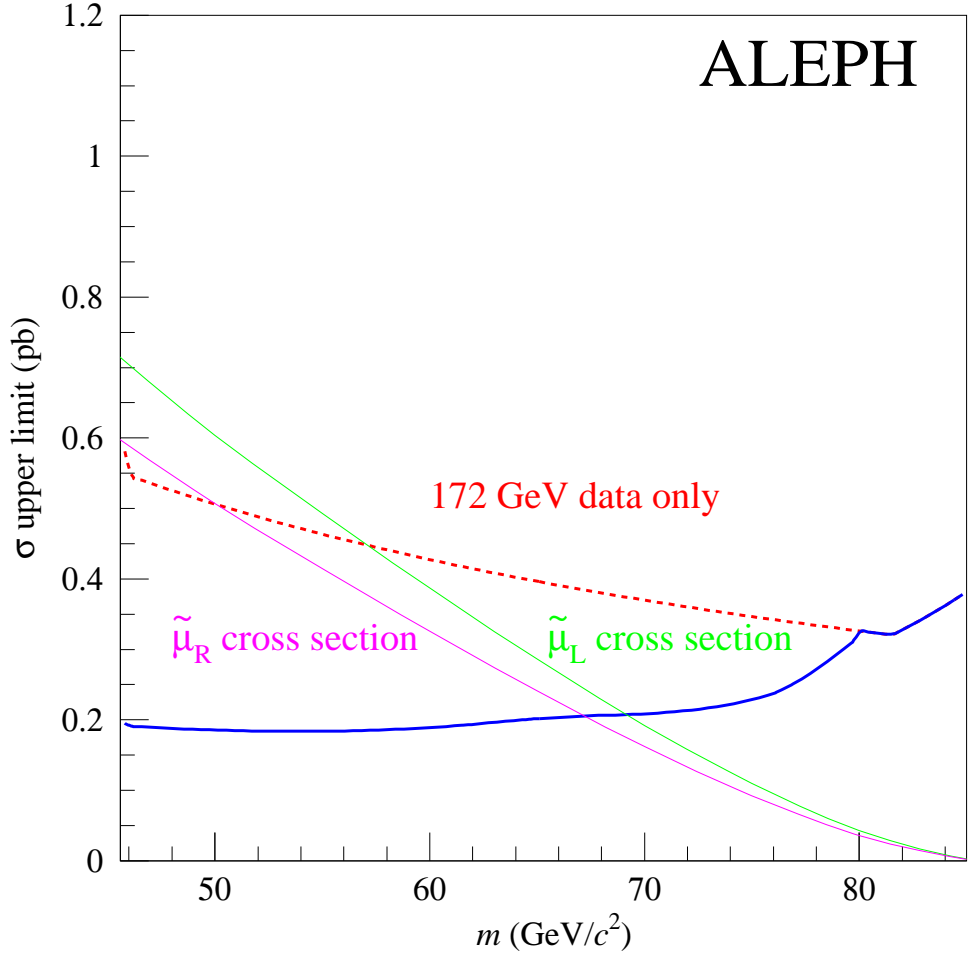,
height=15.cm}}
\end{center}
\caption{95$\%$ CL upper limit on the production cross section
for spin-0 particles at $\sqrt{s}=172$~GeV, as a function of mass.
The dashed curve shows the limit obtained from the 172~GeV data set alone, 
while the thick solid curve shows the limit obtained by
adding the 161, 136 and 130~GeV data to the analysis. The left-handed and
right-handed scalar-muon (or scalar-tau) production cross sections  
are superimposed.}
\label{figura2}
\end{figure}

No event in the 172~GeV data survives any of the three selections.
This sets the 95$\%$ confidence level upper limits on the production 
cross section which are plotted as dashed curves
in Fig.~\ref{figura1} and~\ref{figura2} 
for spin-1/2 and spin-0 particles, respectively.
No candidate is found in the data samples collected at 
$\sqrt{s}= 161$, 136 and 130~GeV, allowing 
the cross section limit to be improved.

The corresponding integrated luminosities must be
rescaled to account for the  dependence on $\sqrt{s}$
of the spin-0 and spin-1/2 production cross sections shown 
in Eq.~\ref{css}.
The combined upper limit at 95$\%$~CL is
$$\sigma_{172}\leq \frac {3} {({\cal L}_{172} \epsilon_{172} +
{\cal L}_{161}^{\mathrm{res}} \epsilon_{161} +
{\cal L}_{136}^{\mathrm{res}} \epsilon_{136} +
{\cal L}_{130}^{\mathrm{res}} \epsilon_{130})}\ ,$$ 
where ${\cal L}^{\mathrm{res}}$ denotes the rescaled luminosities
and $\epsilon$ the selection efficiencies.
The improved limits
obtained by including the data at lower energies are shown as thick curves 
in Fig.~\ref{figura1} and ~\ref{figura2}.
Cross sections of 0.2--0.4~pb at $\sqrt{s} = 172$~GeV are excluded at 95\%~CL
for masses between 45 and 86~GeV/$c^2$, almost independently of
the mass and spin of the  heavy particle.

\begin{figure}
\begin{center}\mbox{\epsfig{file=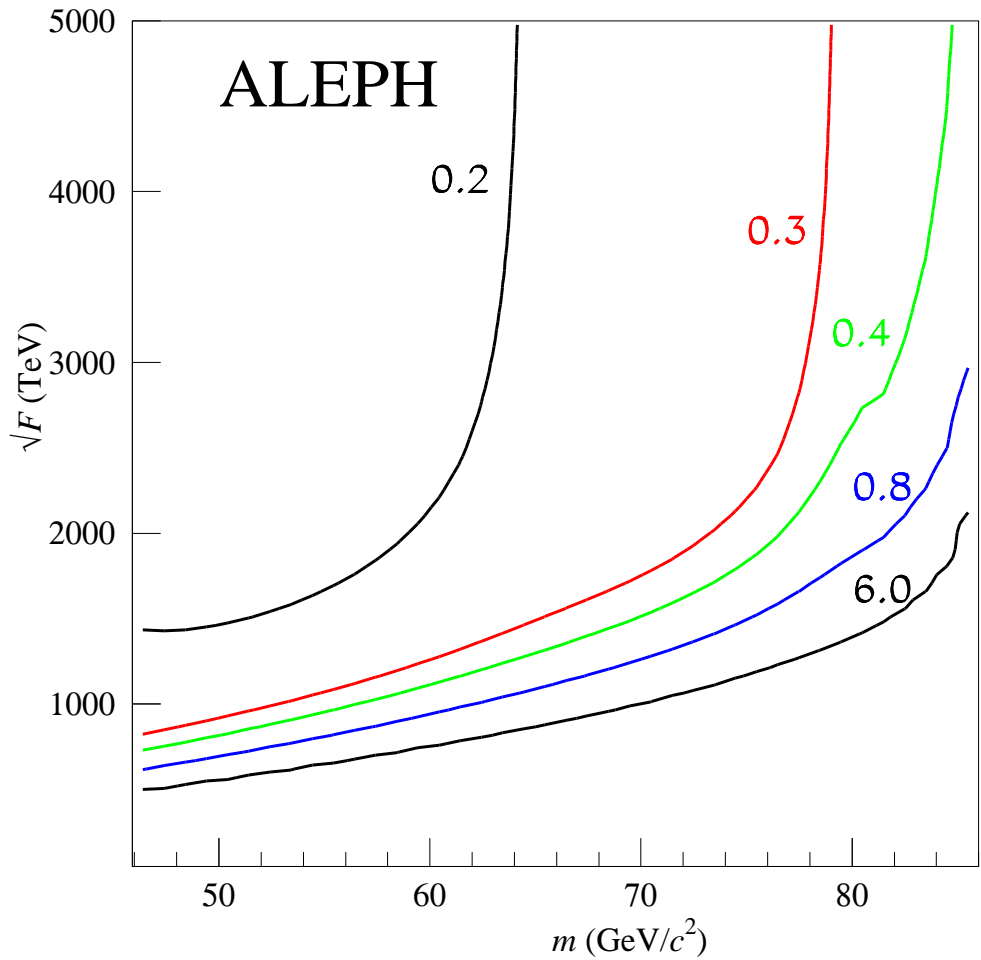,
height=15.cm}}
\end{center}
\caption{95$\%$ CL upper limits (in pb) on the production cross section
for spin-0 particles at $\sqrt{s}=172$~GeV as a function of mass
and SUSY-breaking scale, $\sqrt{F}$.}
\label{figura4}
\end{figure}

As already mentioned, the possibility
of a long-lived charged scalar lepton depends, in the model
of Ref.~\cite{Dimopoulos}, on the value
of the SUSY breaking scale parameter $\sqrt{F}$;
the requirement that the particle decay
outside the detector therefore constrains this scale. In Fig.~\ref{figura4} 
the cross section limit is plotted as a
function of $m$ and $\sqrt{F}$.  The lifetime-dependence of the efficiency is
approximated with a factor
$\exp{(-2 \ell_{\mathrm{det}}/\beta\gamma c \tau)}$, 
where $\ell_{\mathrm{det}}=8.5$~m 
is the maximum length travelled by a particle inside the detector. 
The limit on the cross section worsens for scales below 500--1000~TeV.
 
These cross section upper limits can be translated into 
lower limits on the masses of charginos and sleptons with long lifetimes.
The cross sections expected in the MSSM~\cite{Ridolfi} for charginos 
and for right- and left-handed smuons or staus are also shown
in Fig.~\ref{figura1} and~\ref{figura2}, respectively. 
The selectron production cross section also depends on the neutralino
masses and couplings due to neutralino exchange in the $t$ channel,
therefore no general mass limits can be given.
For charginos the cross section range displayed in Fig.~\ref{figura1} 
was determined as follows:
a value of $\tan{\beta}=\sqrt{2}$ was chosen; 
a given chargino mass then defines a relation between M and $\mu$ 
(where $\mu$ is the supersymmetric mass term
which mixes the two Higgs superfield);
for each pair of $M$ and $\mu$ values, the gauge unification condition was relaxed and 
$M^{'}$ varied up to $5~M$; if the condition $m_{\tilde{\chi}^{\pm}} - 
m_{\tilde{\chi}^{0}} \leq 200$~MeV/$c^2$ could be satisfied in that way, 
the chargino production cross section was calculated, assuming a sneutrino mass of 
250~GeV/$c^2$.

The lower 95$\%$ confidence level 
mass limits for long-lived smuons or staus are
67 and 69~GeV/$c^2$ for right- and left-handed
particles, respectively. For a long-lived chargino, due to the large
production cross section, the kinematic limit of 86~GeV/$c^2$ is almost attained.

\section{Conclusions}

ALEPH data collected at 172, 161, 136 and 130~GeV yield
a 95$\%$ confidence level upper limit of 
0.2--0.4~pb on the pair-production
cross section at 172 ~GeV of long-lived, singly-charged particles with masses 
between 45 and 86~GeV/$c^2$.
This cross section limit implies, in the MSSM,
lower limits of 67 ~GeV/$c^2$ on the mass of right-handed staus 
and smuons and of 69~GeV/$c^2$ on the mass of left-handed staus 
and smuons.
For long-lived charginos a lower mass limit of $\sim 86$~GeV/$c^2$ 
has been obtained.

\section*{Acknowledgements}

We wish to thank our colleagues in the CERN accelerator divisions for the
successful operation of the LEP storage ring at high energy. 
We also thank the engineers and technicians in all our institutions for 
their support in constructing and operating ALEPH. Those of us from non-member states
thank CERN for its hospitality.

\newpage
\setlength{\textwidth}{18.0cm}
\setlength{\textheight}{24.0cm}

\newpage
\setlength{\textwidth}{18.0cm}
\setlength{\textheight}{24.0cm}

\newpage
\setlength{\textwidth}{18.0cm}
\setlength{\textheight}{24.0cm}

\end{document}